\documentclass[letterpaper,10pt,conference]{ieeeconf}
\IEEEoverridecommandlockouts
\usepackage{amsmath,amssymb,mathtools}
\usepackage{bm}
\usepackage{graphicx}
\usepackage{xcolor}
\usepackage{cite}
\usepackage{booktabs}
\usepackage{url}
\usepackage[hidelinks]{hyperref}
\usepackage{caption}
\usepackage{subcaption}
\usepackage{makecell}
\usepackage[normalem]{ulem}
\newcommand{\R}{\mathbb{R}}

\title{Uncertainty-Aware Vision-Based Autonomous Aerial Refueling}
\author{%
Minhyuk Jang$^{1,*}$, Thinh Nguyen$^{1,*}$, Vivek Khatana$^{1,2}$,
Sandeep Banik$^{1}$, Naira Hovakimyan$^{1}$, Petros G. Voulgaris$^{2}$%
\thanks{This work was supported in part by the National Aeronautics and Space Administration (NASA) under Grant 80NSSC22M0070, 
the National Science Foundation (NSF) under Grants CMMI 2135925, CPS 2311085 and IIS 2331878.}
\thanks{$^{*}$Minhyuk Jang and Thinh Nguyen contributed equally.}%
\thanks{$^{1}$Minhyuk Jang, Thinh Nguyen, Vivek Khatana, Sandeep Banik,
and Naira Hovakimyan are with the Department of Mechanical Science and
Engineering, University of Illinois Urbana-Champaign, Urbana, IL-USA.
{\tt\small \{jang64, thinhqn2, vkhatana, baniksan, nhovakim\}@illinois.edu}}%
\thanks{$^{2}$Vivek Khatana and Petros G. Voulgaris are with the Department of Mechanical Engineering,
University of Nevada, Reno, NV-USA.{\tt\small \{vkhatana,pvoulgaris\}@unr.edu}}%
}

\begin{document}
\maketitle
\thispagestyle{empty}
\pagestyle{empty}

\begin{abstract}
Probe-and-drogue refueling (PDR) requires accurate terminal alignment of a receiver probe with an uninstrumented, dynamically moving drogue despite uncertainty in both relative-state estimation and closed-loop motion. Existing vision-based docking approaches provide relative-state estimates, but generally do not quantify estimation and operational uncertainty in a way that informs the control authority required for robust docking performance. We present an uncertainty-aware vision-based PDR docking architecture that fuses keypoint-based drogue measurements, differential GPS, and avionics in an extended Kalman filter. Split conformal calibration characterizes near-contact error in the fused probe-to-drogue displacement estimate, while a covariance-based operational uncertainty quantification designs position-dispersion limits along with corrective-command variance budgets under modeled handoff variability and disturbances. The resulting uncertainty statistics define geometric and command margins for a range-conditioned deterministic predictive controller.
In high-fidelity FlightGear/JSBSim simulations with four controller variants and 50 runs per variant, the combined-margin controller achieved 45/50 centered basket-plane crossings versus 34/50 for nominal MPC. Median planar crossing error decreased from 0.231 to 0.175~$\mathrm{m}$ and the empirical 90th percentile from 0.403 to 0.286~$\mathrm{m}$, while the roll-command RMS increased from 0.22$^\circ$ to 0.48$^\circ$.
\end{abstract}

\section{Introduction}\label{sec:introduction}
Autonomous aerial refueling (AAR) extends aircraft endurance without landing. One approach is probe-and-drogue refueling (PDR), in which the receiver guides its probe into a drogue towed by the tanker. This docking maneuver is particularly challenging because the flexible hose--drogue assembly responds to atmospheric disturbances, tanker wake, and receiver bow-wave effects, while receiver corrections are constrained by aircraft response and command authority~\cite{jinrui2024progress,ren2019reliable}.
Small errors near contact can cause the probe to miss the drogue despite a well-tracked nominal trajectory~\cite{costello2023airworthiness}. 
Terminal docking is not only a perception or tracking problem, but also a constrained decision problem: \textit{how much of the finite capture region should be reserved for estimation error and motion uncertainty, and how much control authority should remain available for corrective response?}

Vision provides close-range measurements without instrumenting the drogue~\cite{martinez2013vision,ma2019monocular}. The docking control system must translate uncertain relative-position estimates into control inputs that achieve alignment despite disturbances and variable approach conditions. We distinguish \emph{estimation uncertainty}, the error in the fused probe-to-drogue vision-aided multisensor estimate, from \emph{operational uncertainty}, the trajectory dispersion induced by variable handoff conditions, disturbances affecting drogue motion, and receiver-response error. A nominal trajectory through the basket center alone does not indicate how much uncertainty can be accommodated or what corrective response is required. We therefore seek a systematic basis for selecting planning margins that reflects both uncertainty and receiver dynamics. To this end, we combine data-driven uncertainty calibration with a model-based feasibility assessment. Conformal prediction converts estimator error into a geometric allowance through calibrated prediction sets, with coverage specified under the calibration assumptions~\cite{vovk2005algorithmic,angelopoulos2024conformal,barber2023conformal}. Covariance uncertainty quantification~\cite{okamoto2018optimal,liu2024optimal,rapakoulias2023dtcs} evaluates whether a desired near-contact dispersion can be supported within control-command budgets under operational uncertainty. Together with the estimator allowance, the resulting position and command uncertainty statistics configure a range-conditioned online model predictive controller (MPC) that generates roll, pitch, and airspeed references for the existing automatic flight control system (AFCS). The major contributions of this work are as follows: 

\textbf{(i)} a two-source uncertainty decomposition for vision-based PDR for AAR, in which conformal calibration maps fused-estimator error to a geometric allowance and offline covariance design maps operational uncertainty to coupled position/command margins; \textbf{(ii)} a computationally light offline--online controller architecture in which these margins configure a deterministic range-conditioned MPC above an existing AFCS; and \textbf{(iii)} a high-fidelity simulation ablation and feasibility study that reveals the tradeoff between retained capture tolerance and control-command variability, while isolating the effects of estimator and operational-uncertainty margins on crossing accuracy and command effort.

Additional videos, implementation details, and supplementary results are available on the project website:
\url{https://aar-refueling.pages.dev/}

\begin{figure*}[t]
\centering
\includegraphics[width=0.9\textwidth]{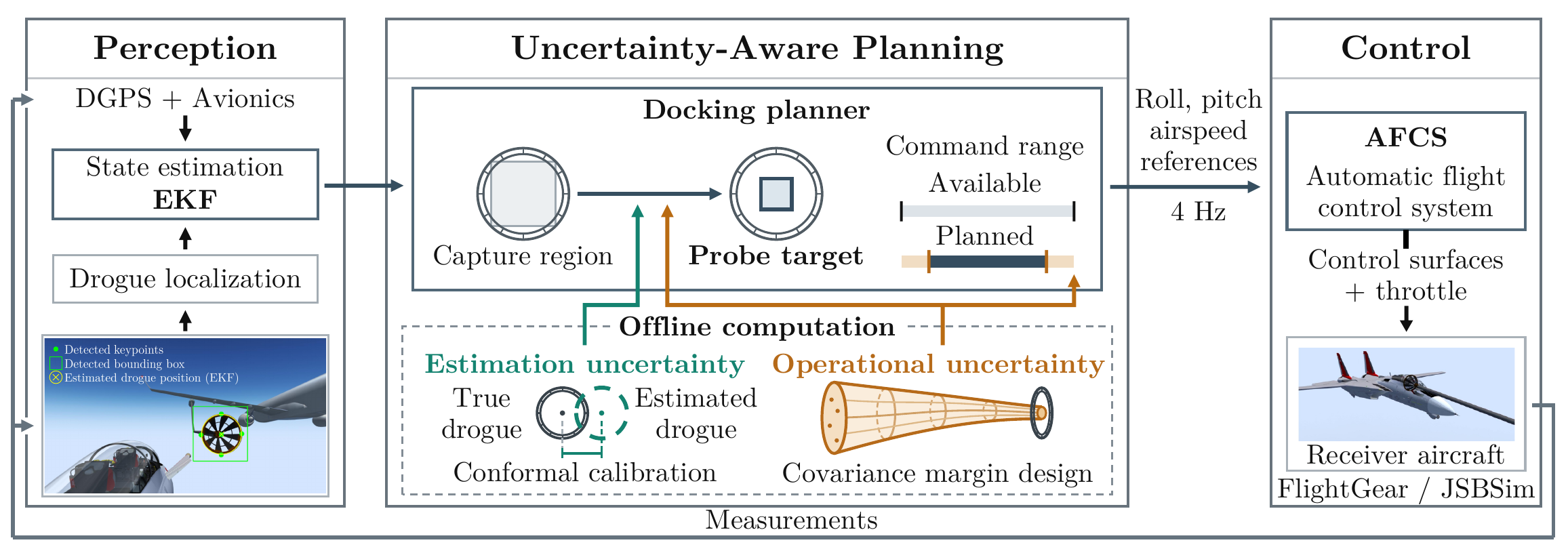}
\caption{\small 
Uncertainty-aware AAR architecture. An EKF fuses vision, DGPS, and avionics. The offline covariance margin design supplies paired geometric and command margins; conformal calibration supplies an estimator allowance. The cone depicts the relative-position spread under the offline affine policy, whose margin schedule configures the online MPC.
The MPC targets 4-Hz updates and supplies roll, pitch, and airspeed references for AFCS tracking.
}
\label{fig:architecture}
\end{figure*}

\section{Related Work}\label{sec:related-work}
\textbf{Relative sensing and estimation.} Refueling navigation must distinguish aircraft-relative motion from motion of the drogue itself. Vision/GPS fusion estimates tanker--receiver relative position~\cite{mammarella2008machine}, while stereo/inertial fusion supports relative navigation without GPS~\cite{johnson2017combining}. Terminal docking additionally requires direct drogue observations. Monocular tracking and structural geometry provide drogue position and pose measurements~\cite{martinez2013vision,ma2019monocular}, while learned keypoints support geometric pose recovery~\cite{wen2025aarpose} and neural perception has been integrated with unscented filtering~\cite{wagner2026towards}. 
Recent work has further improved pose estimation in challenging visual conditions using lightweight learned architectures~\cite{wen2025aarpose}. 
These methods primarily improve the quality and robustness of the visual measurement itself. In contrast, we focus on the residual uncertainty remaining after vision is fused with navigation and avionics, and on converting that fused-estimator error into a geometric allowance for docking.

\textbf{Docking control under uncertainty.}
Vision-based trajectory tracking follows drogue motion using relative measurements~\cite{tandale2006trajectory}.
For uncertain aircraft response, $\mathcal{L}_1$ adaptive flight control provides a priori performance bounds and a time-delay margin bounded under an assumption of known drogue positions; its AAR evaluation includes wake-induced aerodynamic uncertainty and partial loss of control effectiveness~\cite{wang2010verifiable}.
More recent work has derived closed-loop docking-performance bounds under uncertain drogue motion~\cite{Jayarathne} and developed robust PDR docking controllers under nonlinearities, disturbances, and actuator uncertainty~\cite{hang2025robust,ren2025multi}. 
Terminal iterative learning uses earlier docking errors to update initial positions and feedforward commands~\cite{ren2019reliable}, while data-driven safety assessment predicts docking outcomes to guide the approach strategy~\cite{liu2021dockingsafety}.
Our approach is complementary: we connect uncertain docking conditions to the receiver's corrective capability and seek to determine how much geometric tolerance and corrective-command authority should be reserved before online planning to accommodate quantified uncertainty.

\textbf{Uncertainty quantification and planning.} Conformal prediction provides finite-sample marginal coverage under exchangeability~\cite{vovk2005algorithmic,angelopoulos2024conformal}, with extensions that quantify coverage loss when this assumption is violated~\cite{barber2023conformal}. In robotics, calibrated keypoint sets propagate image uncertainty into pose regions~\cite{yang2023object}, while calibrated state-estimation regions support measurement-robust barrier-function control under stochastic sensor noise~\cite{yang2023safeperception}. For dynamic environments, trajectory-prediction regions also support MPC among moving agents~\cite{lindemann2023safe,dixit2023adaptive}. Beyond estimation error, disturbances induce state uncertainty whose evolution depends on the system dynamics. Covariance shaping addresses this dependence by propagating dynamic uncertainty through stochastic models and enforcing state and input constraints under uncertainty~\cite{liu2024optimal}. Risk-aware formulations similarly couple state and input statistics through stochastic dynamics~\cite{okamoto2018optimal,rapakoulias2023dtcs}, including applications to motion planning with collision-avoidance constraints~\cite{DRCS}. Existing approaches therefore provide strong tools for calibrated perception uncertainty, stochastic state propagation, and uncertainty-aware control, but these mechanisms are generally treated separately. In PDR docking, estimation error and dynamics-dependent operational uncertainty consume the finite capture and command budgets in different ways. 
We therefore use conformal calibration to determine the geometric allowance associated with fused-estimator error, and an offline covariance-feasibility design to determine compatible position-dispersion and corrective-command margins under operational uncertainty.
Only these margins enter the online MPC, which remains deterministic.

\section{System Model and Estimation}\label{sec:system_model}
\subsection{Docking Task and Coordinate Frames}\label{subsec:frame}
\begin{figure}[t]
    \centering
    \includegraphics[width=\columnwidth]{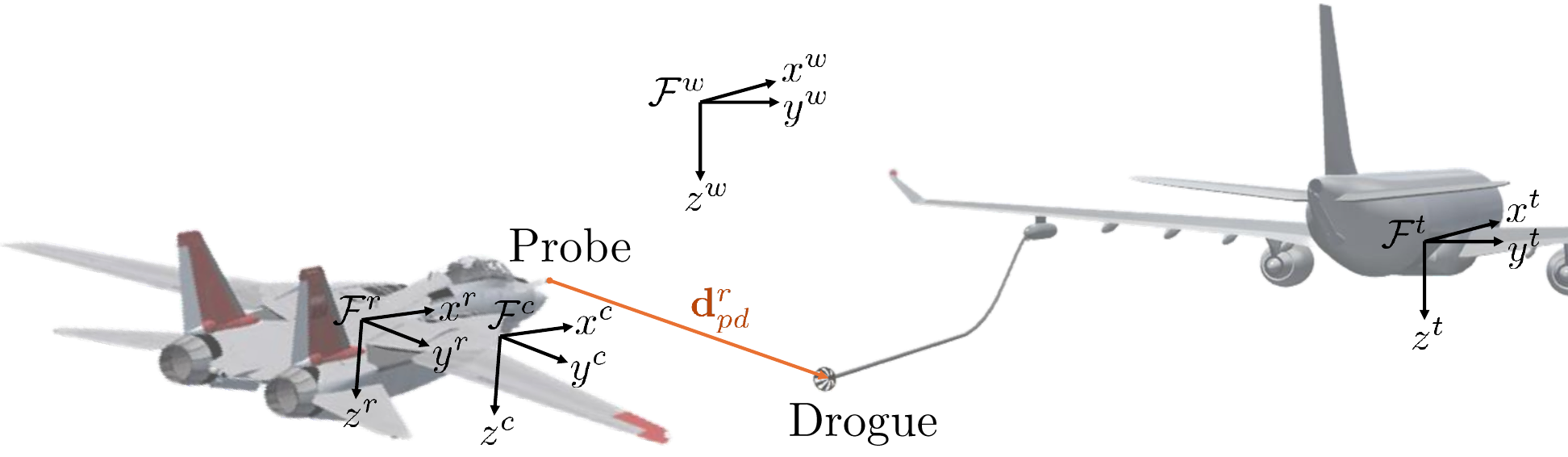}
    \caption{\small Coordinate frames in the aerial refueling system.}
    \label{fig:aar_frames}
\end{figure}

PDR comprises rendezvous and formation joining, pre-contact positioning, docking, fuel transfer, and disconnection~\cite{jinrui2024progress}. We consider terminal PDR docking following a DGPS-guided approach. At docking handoff, vision is activated to observe the uninstrumented drogue, whose tanker-relative motion is not directly measured by DGPS. An extended Kalman filter (EKF) fuses drogue vision, DGPS, and avionics, while the MPC commands roll, pitch, and airspeed references within AFCS limits. We evaluate probe passage through the modeled capture region at the first basket-plane crossing; mechanical engagement and fuel transfer are outside the scope.
Let $\mathcal{F}^w$, $\mathcal{F}^t$, $\mathcal{F}^r$, and $\mathcal{F}^c$ denote the world, tanker, receiver, and camera frames, respectively. We use $\mathbf{p}_a^b$ for the position of point $a$ relative to the origin of $\mathcal{F}^b$, expressed in $\mathcal{F}^b$, $\mathbf{R}_b^c$ for the rotation matrix mapping vectors from $\mathcal{F}^b$ to $\mathcal{F}^c$, and $\mathbf{d}_{ab}^c$ for the displacement vector from point~$a$ to point~$b$, expressed in $\mathcal{F}^c$. Receiver-body axes follow the aerospace convention: $x$ forward, $y$ right, $z$ down. The camera axes are $x$ along the optical axis, $y$-right, $z$-down.  Both frames are illustrated in Fig.~\ref{fig:aar_frames}. For docking alignment, we also define a level approach frame with $x$ parallel to the nominal approach direction and normal to the nominal basket plane, $y$ right, and $z$ down.

\subsection{Relative Dynamics and AFCS Model}
Let $\mathbf p_t^w$ and $\mathbf p_r^w$ denote the tanker and receiver center-of-gravity (CG) in the world frame, and $\mathbf p_d^t$ the drogue position in the tanker frame.
Over the duration of the task,  we assume a steady, straight-and-level tanker track~\cite{costello2023airworthiness}, with known attitude $\mathbf R_t^w$ ($\dot{\mathbf{R}}_t^w = \mathbf{0}$) and airspeed $\mathbf{v}_t^t$ ($\dot{\mathbf{v}}_t^t = \mathbf{0}$).
We adopt a double-integrator model for the drogue motion relative to the tanker
\begin{equation}\label{eq:drogue_dyn}
    \ddot{\mathbf{p}}_d^t = \mathbf{w}_d(t),
\end{equation}
where $\mathbf{w}_d$ is a lumped disturbance acceleration that captures hose dynamics, aerodynamic wake, and atmospheric turbulence. The receiver kinematics depend on its attitude 
$\bm\Theta_r= [\phi_r \ \theta_r \ \psi_r ]^\top$
(roll, pitch, yaw in the ZYX Euler-angle convention), body angular rate $\bm\omega_r \in \mathbb{R}^3$, and body-frame translational velocity $\mathbf{v}_r^r \in \mathbb{R}^3$.
The receiver has two known fixed lever arms, namely, the probe tip $\mathbf p_p^r$ and camera optical center  $\mathbf p_c^r$.
Probe-to-drogue displacement satisfies
$\mathbf d_{pd}^r = (\mathbf R_r^w)^\top
(\mathbf R_t^w\mathbf p_d^t+\mathbf p_t^w-\mathbf p_r^w)-\mathbf p_p^r.$
Differentiating and applying the transport theorem~\cite{Baruh_analytical_1999} yields
\begin{equation}
    \dot{\mathbf{d}}_{pd}^r
      = (\mathbf{R}_r^w)^\top \mathbf{R}_t^w
        \bigl(\dot{\mathbf{p}}_d^t + \mathbf{v}_t^t\bigr)
        - \mathbf{v}_r^r
        - \bm{\omega}_r \times
          \bigl(\mathbf{p}_p^r + \mathbf{d}_{pd}^r\bigr).
    \label{eq:rel_vel}
\end{equation}

For the calm-air (negligible wind) testbed and under the assumption of zero sideslip, we approximate the body-frame velocity as $\mathbf v_r^r= v_r \begin{bmatrix}
    \cos\alpha_r & 0 & \sin\alpha_r
\end{bmatrix}^{\top}$, where $v_r$ is the airspeed and $\alpha_r$ is trim angle of attack.
The AFCS accepts roll, pitch, and airspeed commands (attitude loop).
Its closed-loop response is approximated by first-order lag models for each commanded axis:
\begin{equation}\label{eq:afcs_model} 
    \dot\phi_r \!=\! \frac{\phi_{\mathrm{cmd}}-\phi_r}{\tau_\phi}, \
    \dot\theta_r \!=\! \frac{\theta_{\mathrm{cmd}}-\theta_r}{\tau_\theta}, \
    \dot v_r \!=\! \frac{v_{\mathrm{cmd}}-v_r}{\tau_v},
\end{equation}
with identified time constants $(\tau_\phi,\tau_\theta,\tau_v)$, reported in Section~\ref{sec:experiments}. 
Heading $\psi_r$ is not independently commanded; the AFCS maintains coordinated flight, giving
\begin{equation}
    \dot{\psi}_r = (g/v_r)\tan\phi_r,
    \label{eq:heading_kinematics}
\end{equation}
where $g$ is gravitational acceleration.
The body rate in~\eqref{eq:rel_vel} $\bm{\omega}_r$ is related to the Euler rates $\dot{\bm{\Theta}}_r = [\dot{\phi}_r \ \dot{\theta}_r \ \dot{\psi}_r]^\top$ produced by~\eqref{eq:afcs_model}--\eqref{eq:heading_kinematics} via the  ZYX Euler kinematic relation~\cite{diebel2006representing}:
\begin{equation}
    \bm\omega_r=
    \begin{bmatrix}
        1&0&-\sin\theta_r\\
        0&\cos\phi_r&\sin\phi_r\cos\theta_r\\
        0&-\sin\phi_r&\cos\phi_r\cos\theta_r
    \end{bmatrix}\dot{\bm\Theta}_r.
\label{eq:body_rate}
\end{equation}
Collecting~\eqref{eq:drogue_dyn}--\eqref{eq:body_rate}, we define the state and input vectors as
\begin{equation}
\begin{aligned}
    \mathbf{x} &= [ {\mathbf{p}_d^t}^{\top} \
        \dot{{\mathbf{p}}_d^t}^{\top} \
        {\mathbf d_{pd}^r}^{\top} \
        \bm\Theta_r^\top \
        v_r
    ]^\top\in\R^{13}, \\
    \mathbf{u} &= [
        \phi_{\mathrm{cmd}} \ \theta_{\mathrm{cmd}} \ v_{\mathrm{cmd}}]^\top\in\R^3.
\end{aligned}
\label{eq:state_input}
\end{equation}

Let $\mathbf{f}: \mathbb{R}^{13} \times \mathbb{R}^{3} \to \mathbb{R}^{13}$ denote the deterministic dynamics encompassing~\eqref{eq:drogue_dyn}--\eqref{eq:body_rate}.
Drogue-acceleration disturbances and AFCS roll--pitch response residuals enter through the stochastic model
\begin{equation}\label{eq:nonlinear_model}
\dot{\mathbf x}=\mathbf f(\mathbf x,\mathbf u)+\mathbf G_c\mathbf w_c(t),
\end{equation}
where $\mathbf{w}_c(t) \in \mathbb{R}^{5}$ is a zero-mean white-noise process with spectral-density matrix $\mathbf{Q}_c$, and $\mathbf{G}_{c} \in \mathbb{R}^{13 \times 5}$ is an injection matrix that maps specific noise components to associated state equations.
The vector $\mathbf{w}_c(t)$ decomposes as
\begin{equation}
    \begin{aligned}
        \mathbf w_c(t) &= \begin{bmatrix}
            \mathbf w_d(t)^\top & w_\phi & w_\theta
        \end{bmatrix}^\top\in\R^5, 
    \end{aligned}
\label{eq:noise_injection}
\end{equation}
where $w_\phi$ and $w_\theta$ represent roll- and pitch-tracking residuals. 
The injection matrix is defined as 
$\mathbf G_c =[e_4\ e_5\ e_6 \ e_{10}\ e_{11}],$ where $ e_{i} \in \mathbb{R}^{13}$ is the $i$-th standard basis vector. 
The spectral-density matrix of $\mathbf{w}_c$ is diagonal, reflecting the assumption that the five noise sources are mutually uncorrelated, of the form: $\mathbf Q_c=\operatorname{diag}(q_x,q_y,q_z,\sigma_\phi^2,\sigma_\theta^2)$,
where $q_x,q_y,q_z$ are the acceleration spectral densities, and $\sigma_\phi^2,\sigma_\theta^2$ are AFCS angular-rate residual intensities. 
The numerical value of $\mathbf Q_c$ is identified in Section~\ref{sec:offline_results}.

\subsection{Vision-Based Drogue Detector}\label{sec:ekf}
A receiver-mounted monocular camera uses YOLO11s-pose~\cite{yolo11_ultralytics} to detect the drogue throat center $(u_0,v_0)$ and four evenly spaced rim points,
as shown in Fig.~\ref{fig:architecture}. With known drogue diameter $D$ and calibrated camera intrinsics $(f,c_x,c_y)$, using the pinhole model, the projected ring semi-major axis $a$ gives the approximate depth $Z_d\approx fD/(2a)$.
The resulting camera-frame drogue measurement is
\begin{equation}
    \Tilde{\mathbf{p}}_d^c
      = Z_d
        \begin{bmatrix}
            1 & (u_0 - c_x)/f & (v_0 - c_y)/f
        \end{bmatrix}^{\top},
    \label{eq:vision_measurement}
\end{equation}
where the first component is depth along the optical axis, consistent with
the camera-frame convention. 

\subsection{Multisensor State Estimator}
The estimator fuses three sensor groups, each providing partial observations of the state vector~$\mathbf{x}$:\footnote{Tildes denote measured quantities to distinguish them from ground truth or estimated ones.} (i) DGPS provides the receiver-to-tanker displacement $\Tilde{\mathbf{d}}_{rt}^w \in \mathbb{R}^3$, (ii) Vision gives the drogue position $\Tilde{\mathbf{p}}_d^c \in \mathbb{R}^3$~\eqref{eq:vision_measurement}, and (iii) Avionics provides receiver attitude $\Tilde{\bm{\Theta}}_r \in \mathbb{R}^3$ and airspeed $\Tilde{v}_r \in \mathbb{R}$. 
Physically, DGPS measures the displacement from the receiver CG to the tanker CG in the world frame, expressed as $\mathbf{d}_{rt}^w(\mathbf{x})
      = \mathbf{R}_r^w \bigl(\mathbf{d}_{pd}^r
        + \mathbf{p}_p^r\bigr)
        - \mathbf{R}_t^w\,\mathbf{p}_d^t.$
The vision prediction maps the state to the expected drogue position in the camera frame, expressed as $\mathbf{p}_d^c(\mathbf{x})
      = (\mathbf{R}_c^r)^\top \bigl(\mathbf{d}_{pd}^r
        + \mathbf{p}_p^r
        - \mathbf{p}_c^r\bigr).$
The attitude and airspeed predictions are direct readouts of the corresponding states: $\bm{\Theta}_r(\mathbf{x}) = \bm{\Theta}_r$ and $v_r(\mathbf{x}) = v_r$.
Stacking all three sensor groups yields the composite measurement vector $\mathbf{y} \in \mathbb{R}^{10}$ and measurement function
$\mathbf{h}: \mathbb{R}^{13} \to \mathbb{R}^{10}$:
\begin{equation}
    \mathbf{y}
      \!=\! \mathbf{h}(\mathbf{x}) \!+\! \bm{\nu},
    \
    \mathbf{h}(\mathbf{x}) \!=\! \left[\mathbf{d}_{rt}^w(\mathbf{x})^{\top} \
            \mathbf{p}_d^c(\mathbf{x})^{\top} \ 
            \bm{\Theta}_r^{\top} \
            v_r
            \right]^{\top},
    \label{eq:measurement_model}
\end{equation}
where $\bm{\nu} \in \mathbb{R}^{10}$ is zero-mean measurement noise with covariance $\mathbf{R}_{\nu}$, assumed mutually uncorrelated across sensor groups. 
A continuous-discrete EKF propagates the state and covariance between sensor updates~\cite{lewis1986estimation}. 

\section{Uncertainty-Aware Docking Planning}\label{sec:control}
We separately characterize \textit{estimation} and \textit{operational} uncertainty before incorporating their resulting margins into the PDR docking MPC.
Conformal calibration provides the estimator allowance, while offline covariance design provides compatible position and control-command margins.  

\subsection{Conformal Prediction for Estimation Uncertainty}\label{sec:conformal}

Let $\mathbf d,\widehat{\mathbf d}\in\mathbb R^3$ denote the true and estimated probe-to-drogue displacements in the level approach frame, ordered axial, lateral, and vertical.
We treat the state estimator as a prediction map that produces an estimate $\widehat{\mathbf{d}}_i$ for an input $\boldsymbol{\chi}_i$ (consisting of measurements and process information) \cite{yang2023safeperception}. We use split conformal prediction~\cite{angelopoulos2024conformal,barber2023conformal} to characterize the estimation error from an offline calibration set $\{ (\boldsymbol{\chi}_i, \mathbf{d}_i) \}_{i=1}^{n_c}$. For each $i$, we compute the residual $\mathbf{r}_i := \mathbf{d}_i - \widehat{\mathbf{d}}_i$, with empirical mean $\overline{\mathbf{r}} = \frac{1}{n_c} \sum_{i=1}^{n_c} \mathbf{r}_i$ and covariance $\boldsymbol{\Sigma}_\mathrm{cal} = \frac{1}{n_c-1} \sum_{i=1}^{n_c} (\mathbf{r}_i - \overline{\mathbf{r}}) (\mathbf{r}_i - \overline{\mathbf{r}})^\top$. Using the nonconformity score function $s(\mathbf{d}, \widehat{\mathbf{d}}) := (\mathbf{d} - \widehat{\mathbf{d}})^\top \,\boldsymbol{\Sigma}_\mathrm{cal}^{-1} \,
    (\mathbf{d} - \widehat{\mathbf{d}})$~\cite{mahalanobis1936},
we compute $s_i := s(\mathbf{d}_i, \widehat{\mathbf{d}}_i)$ and sort $s_1, \ldots, s_{n_c}$ in non-decreasing order. For a user-specified error probability $\alpha_c\in(0,1)$, let $\hat{q}$ be the $\lceil(1 - \alpha_{c})(n_c + 1)\rceil$-th element in the sorted list. For a new estimate $\widehat{\mathbf{d}}$, the conformal prediction set given by
\begin{align}\label{eq:conformal_set}
    \mathcal{C}(\widehat{\mathbf{d}}) &= \big\{ \mathbf{d} \in \mathbb{R}^3 ~:~ s(\mathbf{d}, \widehat{\mathbf{d}}) \leq \hat{q} \big\},
\end{align}
satisfies the marginal coverage property \cite[Thm. 2]{barber2023conformal}
\begin{align}\label{eq:conformal_marginal_coverage}
    \hspace{-0.1in} \Pr\big\{ \mathbf{d} \in \mathcal{C}(\widehat{\mathbf{d}}) \big\} \geq
    1 - \alpha_{c} - \textstyle \frac{1}{n_c+1} \sum_{i=1}^{n_c} \mathrm{d}_\mathrm{TV}(Z,Z^i),
\end{align}
where $Z:= \{(\boldsymbol{\chi}_1, \mathbf{d}_1), \ldots, (\boldsymbol{\chi}_{n_c+1}, \mathbf{d}_{n_c+1})\}$. $Z^i$ is $Z$ with the $i$-th and $(n_c+1)$-th elements exchanged, and $\mathrm{d}_\mathrm{TV}$ denotes the total variation distance. When $(\boldsymbol{\chi}_1, \mathbf{d}_1), \ldots, (\boldsymbol{\chi}_{n_c+1}, \mathbf{d}_{n_c+1})$ are exchangeable, the total variation term is zero and the coverage of \eqref{eq:conformal_set} is marginally $1-\alpha_c$.
In practice, exchangeability is often violated and the lower bound in~\eqref{eq:conformal_marginal_coverage} is not computable \cite{angelopoulos2023conformal}.
In Section~\ref{sec:offline_results}, we detail our data collection procedure that maintains a good empirical coverage. To avoid non-affine constraints in the control problem imposed by the ellipsoidal set \eqref{eq:conformal_set}, we consider the bounding box with half-widths
\begin{equation}\label{eq:conformal_margin}
    \rho_x \!=\! \sqrt{\hat{q} [\boldsymbol{\Sigma}_\mathrm{cal}]_{11}},\
    \rho_y \!=\! \sqrt{\hat{q} [\boldsymbol{\Sigma}_\mathrm{cal}]_{22}},\
    \rho_z \!=\! \sqrt{\hat{q} [\boldsymbol{\Sigma}_\mathrm{cal}]_{33}},
\end{equation}
which is more conservative (does not degrade coverage) and introduces affine constraints.

\subsection{Operational Uncertainty Quantification}\label{sec:offline-covariance}

Operational uncertainty arises from variability in the initial state $\mathbf x_0$, including the probe-to-drogue displacement, and disturbances in the receiver--drogue dynamics.
For the offline design, $k=0,\ldots,N$ indexes the states along a nominal trajectory with $N$ discretization intervals.
Linearizing~\eqref{eq:nonlinear_model} about a deterministic nominal trajectory $(\bar{\mathbf x}_k,\bar{\mathbf u}_k)$ and discretizing over an interval $T_c$ gives
\begin{equation}\label{eq:linear_stochastic_model}
\delta\mathbf x_{k+1}=\mathbf A_k\delta\mathbf x_k+\mathbf B_k\delta\mathbf u_k+\mathbf D_k\mathbf w_k,
\end{equation}
where $\mathbf A_k$ and $\mathbf B_k$ are the linearized discrete-time state and input matrices, respectively, $\delta\mathbf x_k=\mathbf x_k-\bar{\mathbf x}_k$, and $\delta\mathbf u_k=\mathbf u_k-\bar{\mathbf u}_k$.
The normalized disturbances $\mathbf w_k$ have zero mean and identity covariance independent across stages and of the initial state $\mathbf x_0$.
We obtain $\mathbf Q_k^d$ by Van Loan discretization~\cite{vanloan1978computing} of the linearized model with state-noise spectral density $\mathbf G_c\mathbf Q_c\mathbf G_c^\top$ over $T_c$, and choose a covariance factor $\mathbf D_k$ satisfying $\mathbf D_k\mathbf D_k^\top=\mathbf Q_k^d$. Let $\mathbf x_h$ denote the full state~\eqref{eq:state_input} at the approach-to-docking handoff, $\bar{\mathbf x}_h := \mathbb{E} [\mathbf x_h] $ and $\bm \Sigma_h := \mathbb{E}[(\mathbf x_h - \bar{\mathbf x}_h)(\mathbf x_h - \bar{\mathbf x}_h)^\top]$. The covariance $\bm \Sigma_h$ captures variability in the pre-contact handoff state, while $\mathbf Q_k^d$ captures uncertainty accumulated during the subsequent docking maneuver. Together, these quantities define the operational uncertainty model. 

The effect of $\bm\Sigma_h$ and $\mathbf Q_k^d$ on docking accuracy depends on the receiver's corrective response. Let $\mathbf e_k:=\delta\mathbf x_k-\mathbb E[\delta\mathbf x_k]$ denote the zero-mean state fluctuation, with covariance $\bm\Sigma_k:=\mathbb E[\mathbf e_k\mathbf e_k^\top]$. For the offline covariance design, we choose the auxiliary feedback policy $\delta\mathbf u_k=\mathbf K_k\mathbf e_k$, so that $\mathbf u_k=\bar{\mathbf u}_k+\mathbf K_k\mathbf e_k$. Substituting this policy into~\eqref{eq:linear_stochastic_model} and subtracting the mean dynamics yields
\begin{equation}
\begin{aligned}
\mathbf e_{k+1}&= (\mathbf A_k +\mathbf B_k \mathbf K_k) \mathbf e_k + \mathbf D_k \mathbf w_k, \\
\bm\Sigma_{k+1}&=(\mathbf A_k+\mathbf B_k\mathbf K_k)\bm\Sigma_k
(\mathbf A_k+\mathbf B_k\mathbf K_k)^\top+\mathbf Q_k^d,
\end{aligned}
\label{eq:mean_covariance_dynamics}
\end{equation}
with $\bm \Sigma_0 = \bm \Sigma_h$. The corresponding control-input covariance 
\begin{align}\label{eq:control_input_covariance}
    \bm \Sigma_{u,k} = \mathbf K_k \bm \Sigma_k \mathbf K_k^\top.
\end{align}
Recursions~\eqref{eq:mean_covariance_dynamics} and~\eqref{eq:control_input_covariance} characterize the coupled evolution of state dispersion and the control input variability under the operational uncertainty model. We measure docking performance using the lateral and vertical components of the probe-to-drogue displacement in the level approach frame in Section~\ref{subsec:frame}. Linearizing the lateral and vertical probe-to-drogue displacement components with respect to $\mathbf x$ at the nominal state $\bar{\mathbf x}_k$ gives $\mathbf H_k\in\mathbb R^{2\times13}$, which maps small state deviations to changes in alignment. 
The corresponding alignment covariance is $\bm \Sigma_{d,k} := \mathbf H_k \bm\Sigma_k\mathbf H_k^\top$, and the lateral and vertical standard deviations are $\sigma_{d,y,k} := \sqrt{[\bm \Sigma_{d,k}]_{11}}, \sigma_{d,z,k} := \sqrt{[\bm \Sigma_{d,k}]_{22}} $. Similarly, the control input covariance $\bm \Sigma_{u,k}$ yields the stagewise standard deviations $\sigma_{u,i,k} := \sqrt{[\bm \Sigma_{u,k}]_{ii}}, \ i \in \{1,2,3\}$. 

Let $\bar\sigma_{d,y}$ and $\bar\sigma_{d,z}$ denote prescribed upper bounds on the lateral and vertical alignment standard deviations during the near-contact stages $\mathcal K_c=\{N-3,\ldots,N\}$, and for each command channel $i$, let $\bar{\sigma}_{u,i,k}$ denote the maximum allowable standard deviation of the control input required at time $k$.
The design therefore seeks matrices $\{\mathbf K_k\}$ such that
\begin{align}\label{eq:design_requirement1}
    \sigma_{d,y,k} \leq \bar\sigma_{d,y} \quad \sigma_{d,z,k} \leq \bar\sigma_{d,z} \quad \mbox{for all } k \in \mathcal{K}_c,
\end{align}
while simultaneously satisfying
\begin{align}\label{eq:design_requirement2}
    \sigma_{u,i,k}\leq\bar{\sigma}_{u,i,k},
\quad i\in\{1,2,3\},\quad k=0,\ldots,N-1.
\end{align}

To enforce these coupled position and command-dispersion requirements tractably, we use the covariance-variable formulation of~\cite{rapakoulias2023dtcs}. Let symmetric matrices $\mathbf Z_k \succeq 0$ and $\mathbf Y_k \succeq 0$ be upper bounds on the state and control covariances, respectively, together with an auxiliary variable $\mathbf U_k$ satisfying $\mathbf U_k = \mathbf K_k \mathbf Z_k$. Defining $\mathbf M_k=\mathbf A_k\mathbf Z_k+\mathbf B_k\mathbf U_k$, the covariance propagation and command-dispersion constraints can be written as linear matrix inequalities. For state weights $\mathbf Q_{\Sigma,k}\succeq0$ and command
weights $\mathbf R_{\Sigma,k}\succ0$, we solve
\begin{align}
\min_{\{\mathbf Z_k,\mathbf U_k,\mathbf Y_k\}} &
\sum_{k=0}^{N-1}\left[
\operatorname{tr}(\mathbf Q_{\Sigma,k}\mathbf Z_k)
+\operatorname{tr}(\mathbf R_{\Sigma,k}\mathbf Y_k)\right]
+\operatorname{tr}(\mathbf Q_{\Sigma,N}\mathbf Z_N)
\nonumber\\
&\hspace{-0.6in} \mathrm{s.t.} \ \mathbf Z_0=\bm\Sigma_h, 
\begin{bmatrix}
\mathbf Z_{k+1}-\mathbf Q_k^d&\mathbf M_k \nonumber \\
\mathbf M_k^\top&\mathbf Z_k
\end{bmatrix}\succeq0, \ \begin{bmatrix}
\mathbf Z_k&\mathbf U_k^\top \\
\mathbf U_k&\mathbf Y_k
\end{bmatrix}\succeq0, \\
&[\mathbf H_k \mathbf Z_k \mathbf H_k^\top]_{11}\le\bar\sigma_{d,y}^2,
\quad k\in\mathcal K_c, \label{eq:offline_covariance_problem} \\
&[\mathbf H_k \mathbf Z_k \mathbf H_k^\top]_{22}\le\bar\sigma_{d,z}^2,
\quad k\in\mathcal K_c, \nonumber  \\
&[\mathbf Y_k]_{ii}\le\bar\sigma_{u,i,k}^2, \quad i \in \{1,2,3\}. \nonumber 
\end{align}
The solution yields $ \mathbf K_k := \mathbf U_k \mathbf Z_k^{\dagger}$, where $(\cdot)^{\dagger}$ denotes the Moore–Penrose pseudoinverse. Propagating~\eqref{eq:mean_covariance_dynamics} and~\eqref{eq:control_input_covariance} with these gains gives lateral and vertical alignment standard deviations, together with the stagewise command standard deviations that are retained as the operational-uncertainty statistics passed to the docking online controller. The auxiliary gains $\mathbf K_k$ are not applied online.

\subsection{Uncertainty-Aware Docking MPC}\label{sec:online-mean}

The online MPC uses the estimator allowance $\rho_j$, terminal alignment SDs $\sigma_{d,j}:=\sigma_{d,j,N}$ for $j\in\{y,z\}$, and stagewise control input SDs $\sigma_{u,i,k}$ obtained in Sections~\ref{sec:conformal} and~\ref{sec:offline-covariance} to tighten an otherwise nominal docking problem. 

Here $t$ indexes MPC updates, $l$ indexes prediction stages starting from zero at each update, and $k$ indexes the offline covariance propagation. We omit the update index on quantities defined within the current horizon. At MPC update $t$, a straight-in reference is initialized at the current estimated probe-to-drogue range $\hat r_t = \|\hat{\mathbf d}_t\|$, with zero lateral and vertical offsets in the level approach frame. The range $\hat r_t$ selects an offline stage $s$ and a horizon
of $N_t=N-s$ prediction intervals, where $N$ is the offline covariance propagation horizon length. The reference trajectory $(\bar{\mathbf x}_l,\bar{\mathbf u}_l)$ is generated by propagating the deterministic model under constant trim commands. The MPC optimizes deviations from this reference according to $\bm\mu_{l+1}=\mathbf A_l\bm\mu_l+\mathbf B_l\mathbf v_l,$
where $\bm \mu_l$ denotes the nominal state correction and $\mathbf v_l$ the corresponding control input correction required. 
Let $\mathbf H_{N_t}$ denote the stored alignment Jacobian evaluated offline at the terminal reference state. The predicted terminal alignment is $\mathbf d_{N_t}^{\mathrm{pred}}=\mathbf H_{N_t}\bm\mu_{N_t}$, and the predicted control inputs are $\mathbf u_l^{\mathrm{pred}}=\bar{\mathbf u}_l+\mathbf v_l$
for $l=0,\ldots,N_t-1$.
The terminal capture constraint is tightened to account for both uncertainty sources. Let $b_g$ denote the admissible capture half-width along
each alignment axis. 
Reserving the estimator allowance
$\rho_j$ from~\eqref{eq:conformal_margin} and the operational margin $\kappa\sigma_{d,j}$, with chosen scale $\kappa>0$, leaves
\begin{align}\label{eq:tightened_box}
    \beta_j=b_g-\rho_j-\kappa\sigma_{d,j,N}\ge0, \quad j \in \{y,z\}.
\end{align}
\textit{Thus, to be robust against the estimation and operational uncertainties, the nominal terminal trajectory is required to lie inside a reduced capture region whose size reflects both estimation error and modeled closed-loop dispersion.}

The operational uncertainty propagation procedure also determines how much of the AFCS command authority must be reserved for corrective maneuvers. For each command input channel $i \in \{1,2,3\}$, the lower and upper bounds $u_{i,\min}$ and $u_{i,\max}$ are tightened, respectively, according to
\begin{align}\label{eq:command_bounds}
    \hspace{-0.08in} \underline u_{i,l} = u_{i,\min} + \kappa\sigma_{u,i,s+l}, \ \overline u_{i,l} = u_{i,\max} - \kappa\sigma_{u,i,s+l}.
\end{align}
The geometric and control input margins are coupled through the uncertainty propagation design: achieving smaller near-contact position dispersion can require larger corrective-command variability and reduce the control authority.

For positive-semidefinite state weights $\mathbf Q_\mu,\mathbf Q_{\mu,N}$ and positive-definite command weight $\mathbf R_\mu $, define
\begin{equation*}
\textstyle J_\mu := \sum_{l=0}^{N_t-1}
\left(\bm\mu_l^\top\mathbf Q_\mu\bm\mu_l
+\mathbf v_l^\top\mathbf R_\mu\mathbf v_l\right)
+\bm\mu_{N_t}^\top\mathbf Q_{\mu,N}\bm\mu_{N_t}.
\end{equation*}
Let $\hat{\mathbf x}_t$ denote the current EKF state estimate, $u_{i,t-1}$ the previously applied command in channel $i$, $\dot u_{i,\max}$ its slew-rate limit, and $\mathbf w_\xi>0$ the slack-penalty weights. The online docking control problem is then
\begin{equation}
\begin{aligned}
\hspace{-0.1in}
\min_{\{\bm\mu_l,\mathbf v_l\},\,\bm\xi\ge0}\quad
&J_\mu+\mathbf w_\xi^\top\bm\xi\\
\mathrm{s.t.}\quad
&[\bm\mu_0]_{1:6}=\mathbf0,\
[\bm\mu_0]_{7:13}=[\hat{\mathbf x}_t-\bar{\mathbf x}_0]_{7:13},\\
&\bm\mu_{l+1}=\mathbf A_l\bm\mu_l+\mathbf B_l\mathbf v_l,\\
&\hspace{-0.2in}
|[{\mathbf d}_{N_t}^{\mathrm{pred}}]_1|\le\beta_y+\xi_y,\
|[{\mathbf d}_{N_t}^{\mathrm{pred}}]_2|\le\beta_z+\xi_z,\\
&\underline u_{i,l}\le[\mathbf u_l^{\mathrm{pred}}]_i
\le\overline u_{i,l},\\
&|[\mathbf u_0^{\mathrm{pred}}]_i-u_{i,t-1}|
\le\dot u_{i,\max}T_s+\xi_{u,i},\\
&|[\mathbf u_l^{\mathrm{pred}}
-\mathbf u_{l-1}^{\mathrm{pred}}]_i|
\le\dot u_{i,\max}T_c.
\end{aligned}
\label{eq:online_qp}
\end{equation}
Here $T_s$ is the target controller-update period and $T_c$ is the prediction-stage duration. Dynamics and command bounds apply for $l=0,\ldots,N_t-1$. The final slew constraint
applies for $l=1,\ldots,N_t-1$; terminal alignment is enforced at the horizon endpoint. The slack variables $\bm \xi =[\xi_y,\xi_z,\xi_{u,1},\xi_{u,2},\xi_{u,3}]^\top$
soften the terminal alignment and first-command slew
constraints. At each update, the AFCS receives the control input $\mathbf u_t=\bar{\mathbf u}_{0}+\mathbf v_0^\star$, where $\mathbf v_0^\star$ is the first MPC-optimized command correction.

\section{Experimental Evaluation}\label{sec:experiments}

We first assess estimator-error calibration (Section~\ref{sec:conformal}) and operational-uncertainty statistics via covariance design (Section~\ref{sec:offline-covariance}). 
We examine covariance-margin feasibility and evaluate closed-loop docking performance in high-fidelity simulator.

\subsection{Testbed and Implementation}
\label{sec:testbed}
FlightGear 2024.1 and JSBSim~\cite{berndt2004jsbsim} simulate an F-14B receiver approaching an A330 MRTT tanker flying straight and level at FL220 and 183~m/s in calm air.\footnote{Detection uses Ultralytics 8.4 and PyTorch 2.5, and the online QP~\eqref{eq:online_qp} is solved using Clarabel 0.11~\cite{goulart2026clarabel}. Experiments were run on an Intel Core i7-14700F with 32~GB RAM and an NVIDIA RTX~4060~Ti (8~GB).}
The nominal 20-ft (6.1-m) pre-contact handoff is consistent with the NASA/DARPA flight tests~\cite{dibley2007autonomous}.
The drogue follows lightly damped pendulum dynamics under stationary, constant-intensity Ornstein--Uhlenbeck forcing~\cite{gillespie1996ornstein}. A receiver-mounted camera feeds YOLO11s-pose~\cite{yolo11_ultralytics}, fine-tuned on synthetic images to predict five keypoints. Rendering varies camera-to-drogue distance, lateral and vertical offsets, and receiver attitude; training augmentation includes flipping, translation, scaling, and color variation. Table~\ref{tab:controller_settings} lists the parameters used in the experiments. AFCS time constants are identified by output-error fitting of~\eqref{eq:afcs_model} to data from a dedicated open-loop excitation flight. 

\begin{table}[t]
\centering
\caption{\small Testbed, perception, controller, and covariance uncertainty quantification design settings. Command tuples follow roll, pitch, and airspeed. Unlisted entries and stages of $\mathbf Q_{\Sigma,k}$ are zero.}
\label{tab:controller_settings}
\small
\setlength{\tabcolsep}{3pt}
\renewcommand{\arraystretch}{1.0}
\begin{tabular}{@{}ll@{}}
\toprule
Parameter & Value \\
\midrule
Ring diameter; damping ratio
    & $1.1$~m; $0.2$ \\
Natural frequency (lateral/vertical)
    & $1.0$~rad/s \\
Forcing time constant
    & $3.0$~s \\
Realized drogue RMS (per axis)
    & $0.10$~m \\
\midrule
Image size; frame rate
    & $1848\times1016$~px; $30$~Hz \\
$f$; $(c_x,c_y)$
    & $1616$~px; $(924,508)$~px \\
Images (training/validation/test)
    & $1521/189/114$ \\
Input size; maximum rendered range
    & $1920$~px; $130$~m \\
\midrule
$(\tau_\phi,\tau_\theta,\tau_v)$
    & $(1.358,1.301,1.264)$~s \\
MPC target $T_s$; prediction step $T_c$
    & $0.25$~s; $0.75$~s \\
$\mathbf Q_\mu,\ \mathbf Q_{\mu,N}$
    & $\mathbf I$ \\
$\mathbf R_\mu,\ \mathbf R_{\Sigma,k}$
    & $\operatorname{diag}(3000,10^5,1)$ \\
Command authority about trim
    & $(10^\circ,2.5^\circ,5.0~\mathrm{m/s})$ \\
Slew increments $\dot u_{i,\max}T_c$
    & $(0.57^\circ,0.29^\circ,0.30~\mathrm{m/s})$ \\
$\mathbf w_\xi$
    & $[10^5,10^5,10^4,10^4,10^4]^\top$ \\
$[\mathbf Q_{\Sigma,k}]_{7:9,7:9},\ k\in\mathcal K_c$
    & $\operatorname{diag}(0.3,1,1)$ \\
Alignment SD caps $\bar\sigma_{d,y},\bar\sigma_{d,z}$
    & $b_g/3\simeq0.130$~m \\
Roll-command SD limit $\bar\sigma_{u,\phi}$
    & $3.0^\circ$ \\
Margin scaling
    & $\kappa\simeq 2.06$ \\
\bottomrule
\end{tabular}
\end{table}

\subsection{Uncertainty Characterization}
\label{sec:offline_results}
\noindent\textbf{Conformal Prediction Results.}
We collect calibration data from $7$ independent PDR runs using a baseline LQR controller \cite{kalman1960lqr} and retain samples with depth $\leq 1$ m to reflect the terminal estimator margin~\eqref{eq:tightened_box}.
$10\%$ uniformly random subsampling is performed on each run.
Exchangeability condition is not met: calibration data points from the same run are not independent, and the deployed controller and flight conditions differ. However, our procedure keeps the total variation term in \eqref{eq:conformal_marginal_coverage} small in two ways.
First, we retain the distribution of online PDR operations as opposed to grid sampling \cite{yang2023object} that could cause distribution shift.
Second, random subsampling spaces the samples apart in time and mixes flight conditions, weakening the temporal correlation that causes dependence and nonexchangeability.

We use a calibration dataset with $n_c=481$ and target error rate $\alpha_c=0.1$. On one held-out closed-loop flight, $716/790$ evaluated samples were contained, corresponding to $90.6\%$ empirical pointwise coverage that meets the target $1-\alpha_c=90\%$. The calibrated estimator error margin is $(\rho_x, \rho_y, \rho_z) = (0.082, 0.118, 0.093)$ m.

\noindent\textbf{Covariance Design Results.}
The operational-uncertainty inputs are identified from preceding runs.
The full handoff covariance $\bm\Sigma_h$ is estimated from 100 handoff state vectors; its lateral and vertical relative-position marginal standard deviations are 0.55 and 0.23~m, respectively. Drogue acceleration intensities $(0.0004,0.0104,0.0109)~\mathrm{m^2/s^3}$ are estimated from 125 docking trials. Receiver residuals from 34 runs give roll and pitch diffusion amplitudes of 0.050 and 0.066~deg$\,\mathrm{s}^{-1/2}$, respectively. These amplitudes are converted to radians and squared to form covariance intensities, then discretized using the Van Loan method~\cite{vanloan1978computing}.

The alignment caps $(\bar\sigma_{d,y},\bar\sigma_{d,z})$ and roll-command SD limit $\bar\sigma_{u,\phi}$ in Table~\ref{tab:controller_settings} define the nominal configuration for the margin-allocation study. Covariance propagation verifies both position requirements and gives $(\sigma_{d,y},\sigma_{d,z})\simeq(0.12962,0.12962)$~m and $\max_k\sigma_{u,1,k}\simeq2.424^\circ$. With $(\rho_y,\rho_z)=(0.118,0.093)$~m, applying the selected margins in~\eqref{eq:tightened_box} gives terminal half-widths $(\beta_y,\beta_z)$ of approximately $(0.3889,0.3889)$~m without margins, $(0.2706,0.2964)$~m with the estimator allowance only, $(0.1215,0.1215)$~m with covariance margins only, and $(0.0032,0.0290)$~m with both.
The stagewise command standard deviations additionally determine the command margins in~\eqref{eq:command_bounds} for the covariance-margin and both-margin variants.

\begin{figure}[t]
\centering
\includegraphics[width=\columnwidth]{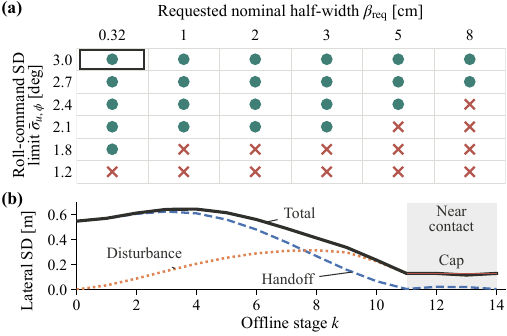}
\caption{\small (a) Compatible ($\bullet$) and incompatible ($\times$) margin allocations; the outline marks the main design.
(b) Lateral SD and its handoff and disturbance contributions under the corresponding auxiliary policy, with the cap applied over $\mathcal K_c$ (shaded).}
\label{fig:certification}
\end{figure}

\noindent\textbf{Margin Feasibility.}
Next, to determine desired nominal capture allowance, we start with each lateral position cap expressed as a requested nominal half-width $\beta_{\mathrm{req}}=b_g-\rho_y-\kappa\bar\sigma_{d,y}$; a feasible policy leaves $\beta_y\ge\beta_{\mathrm{req}}$ while satisfying the command-variance limits. We examine how the lateral position requirement and roll-command variability limit constrain one another in the offline design.
Figure~\ref{fig:certification}(a) re-solves~\eqref{eq:offline_covariance_problem} across these limits, keeping the estimator allowance, uncertainty inputs, objective, and remaining constraints fixed. At a $2.1^\circ$ roll-command SD limit, a 3-cm request is compatible whereas a 5-cm request is not; the latter becomes compatible at $2.4^\circ$.\footnote{We solve~\eqref{eq:offline_covariance_problem} with SD limits reduced by $0.01\%$ to allow for numerical error. Compatible pairs satisfy the original limits after covariance propagation; incompatible pairs have verified infeasibility certificates for the tightened limits.}
The outlined cell marks the main configuration evaluated in the docking. Figure~\ref{fig:certification}(b) separates the handoff and disturbance contributions under the main-design gains.
We propagate the handoff covariance without process noise and the disturbance contribution from zero initial covariance; their covariances sum to the total. The handoff contribution decays under the auxiliary closed-loop control, whereas disturbance-driven variance accumulates along the approach and dominates near contact.\footnote{The same feasibility and covariance-decomposition analyses can be applied to vertical alignment and pitch-command variability.}

\subsection{Closed-Loop Docking Evaluation}
\label{sec:closed_loop_results}
\label{sec:protocol}
Each trial includes the DGPS-guided approach and transition to vision-aided docking, so the MPC starts from the handoff state delivered by the approach controller.

Runs begin 90--110~m behind the tanker, with offsets up to $\pm15$~m laterally and $\pm8$~m vertically.
Handoff requires remaining within $\pm2.0$~m axially of the nominal 20-ft station with closing speed above 0.1~m/s for 1~s.
Station-target offsets are sampled uniformly within $\pm0.8$~m laterally and $\pm0.4$~m vertically, yielding actual handoff offsets up to 1.5 and 0.6~m, respectively.
Docking terminates if range exceeds its handoff value by more than 3~m for over 2~s.

We evaluate four controller variants with 50 runs per variant on identical target-offset and drogue-disturbance seed pairs.
All variants share the estimator, dynamics, costs, reference corrections, and physical command limits.
\emph{Nominal MPC} uses no uncertainty margins; \emph{estimator margin} adds only $\rho$; \emph{covariance margins} add $\kappa\sigma_d$ and the command margins; and \emph{both margins} is the combined controller.
The corresponding terminal half-widths are reported in Section~\ref{sec:offline_results}. At the first basket-plane crossing, let $d_{y,c}$ and $d_{z,c}$ denote the true lateral and vertical probe-to-drogue displacements in the level approach frame. The planar crossing error is $e_c=\sqrt{d_{y,c}^2+d_{z,c}^2}$.
A crossing is \emph{centered} for $e_c\le0.30$~m and lies in the \emph{outer annulus} for $0.30<e_c\le0.55$~m. 

\begin{table}[t]
\caption{Crossing accuracy and roll-command effort (50 runs per variant). P90 is the empirical 90th percentile.}
\label{tab:results}
\centering
\small
\setlength{\tabcolsep}{2pt}
\renewcommand{\arraystretch}{1.08}
\begin{tabular*}{\columnwidth}{@{\extracolsep{\fill}}l c r r r r@{}}
\toprule
Variant & Centered
& \multicolumn{1}{c}{\makecell{Median\\$e_c$ [m]}}
& \multicolumn{1}{c}{\makecell{P90\\$e_c$ [m]}}
& \multicolumn{1}{c}{\makecell{Lateral\\bias [m]}}
& \multicolumn{1}{c@{}}{\makecell{Roll RMS\\{[deg]}}}\\
\midrule
Nominal MPC & 34/50 & 0.231 & 0.403 & $-$0.180 & \textbf{0.22}\\
Estimator margin & 38/50 & 0.230 & 0.373 & $-$0.141 & 0.25\\
Covariance margins & 43/50 & 0.176 & 0.304 & $-$0.042 & 0.33\\
Both margins & \textbf{45/50} & \textbf{0.175} & \textbf{0.286} & \textbf{+0.014} & 0.48\\
\bottomrule
\end{tabular*}
\end{table}
\begin{figure}[t]
\centering
\includegraphics[width=\columnwidth]{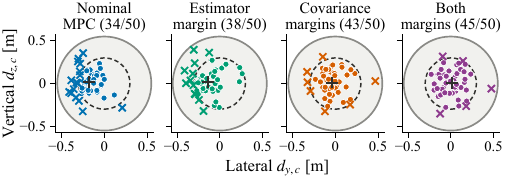}
\caption{\small Main-campaign basket-plane crossings, with 50 runs per variant. The solid 0.55-m circle marks the rim, and the dashed 0.30-m circle marks the centered criterion. Dots and crosses denote centered and outer-annulus crossings, and $+$ marks the mean.}
\label{fig:contacts}
\end{figure}

\noindent\textbf{Crossing Accuracy and Command Activity.}
With tracking costs fixed, MPC with both margins achieved the highest centered-crossing count and lowest median and P90 error among the four variants (Table~\ref{tab:results} and Fig.~\ref{fig:contacts}). Covariance-derived margins account for most of the improvement in median crossing error, reducing it from $0.231$~m to $0.176$~m. Adding the conformal estimator allowance changes the median only marginally, but further reduces the empirical P90 from $0.304$~m to $0.286$~m and increases centered crossings from $43/50$ to $45/50$. This suggests that the operational-uncertainty margins primarily improve nominal centering, while the estimator allowance provides additional robustness in the upper tail. The combined controller also exhibits substantially smaller empirical lateral bias: the mean lateral crossing error shifted from $-0.180$~m for nominal MPC to $0.014$~m with both margins. This improvement came with greater roll-command RMS, increasing from $0.22^\circ$ for nominal MPC to $0.48^\circ$ with both margins.

\begin{figure}[h]
\centering
\includegraphics[width=\columnwidth]{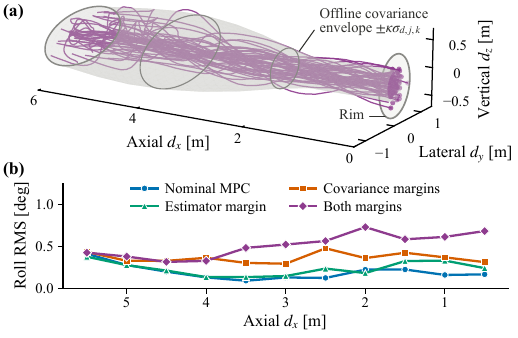}
\caption{\small (a) Approach trajectories with margins overlaid on the offline covariance envelope. The envelope visualizes the offline covariance design and is not a probabilistic tube for the executed MPC trajectories. (b) Roll-command RMS relative to trim, binned by axial distance for all four variants.}
\label{fig:interface}
\end{figure}

Figure~\ref{fig:interface}(a) shows trajectories with both margins on the offline covariance envelope, showing convergence toward the basket from varied handoff offsets. Figure~\ref{fig:interface}(b) shows similar roll activity near handoff but sustained higher activity with both margins near contact, indicating that the additional effort is concentrated in final alignment.

\noindent\textbf{Runtime and Solver Outcomes.}
Offline covariance uncertainty quantification required 5.3~s on average, while online MPC, including QP construction and solution, took a median of 24~ms and a maximum of 126~ms, below the 250~ms target update period.
All online QPs returned an optimal status, and all logged terminal alignment slacks were zero.

\section{Conclusion and Future Work}
We presented a vision-based docking architecture that translates estimation and operational uncertainty into geometric and control input margins. Conformal calibration provides the estimator allowance, while covariance uncertainty quantification provides the position dispersion along with the control input margins required to counter the operational uncertainty. This provides a principled basis for incorporating uncertainty bounds and corresponding control input margins while retaining a deterministic MPC. The high-fidelity experiments demonstrate that the proposed controller achieved 45/50 centered crossings versus 34/50 for nominal MPC.

Future work will extend conformal calibration to trajectory-wide uncertainty tubes for planning within a corridor of autonomy~\cite{costello2023airworthiness}.
and integrate $\mathcal{L}_1$ adaptive control~\cite{hovakimyan2010l1} for improved transient response and disturbance rejection.

\section*{{Acknowledgment}}

The authors used Claude Opus 5 for code implementation assistance and OpenAI GPT-6 for manuscript editing. All research design, experiments, validation, and interpretation were independently developed and verified by the authors.

\bibliographystyle{IEEEtran}
\bibliography{My_Refs}
\end{document}